\documentstyle[12pt,psfig,preprint,aps]{revtex}

\begin{document}

\draft

\title{Strangeness Enhancement in Heavy Ion Collisions -- Evidence for Quark-Gluon-Matter ? 
\footnote{supported by DFG, BMBF, Graduiertenkolleg 
'Theoretische und Experimentelle Schwerionenphysik', Josef~Buchmann Stiftung,
and Alexander~von~Humboldt-Stiftung. S.~A.~B. is supported in part by DOE grant DE-FG02-96ER40945 }}
\author{S.~Soff$^{a,b}$, S.A.~Bass$^{c}$, M.~Bleicher$^{a}$, L.~Bravina$^{d}$, M.~Gorenstein$^{a,e}$, 
E.Zabrodin$^{d}$, H.~St\"ocker$^{a}$, W.~Greiner$^{a}$} 
\address{
        $^a$Institut f\"ur Theoretische Physik der J. W. Goethe-Universit\"at\\
        $^{~}$Postfach 11 19 32, D-60054 Frankfurt am Main, Germany\\
        $^b$Gesellschaft f\"ur Schwerionenforschung, Postfach 110552, D-64220 Darmstadt, Germany\\
        $^c$Department of Physics, Duke University, Durham, NC 27708, USA\\
        $^d$Institut f\"ur Theoretische Physik, Universit\"at T\"ubingen\\
        $^{~}$Auf der Morgenstelle 14, D-72076 T\"ubingen\\
        $^e$Bogolyubov Institute for Theoretical Physics, Kiev, Ukraine\\
}

\maketitle

\begin{abstract}

The centrality dependence of (multi-)strange hadron 
abundances is studied for Pb(158\,A\,GeV)Pb reactions and compared to p(158\,GeV)Pb 
collisions. 
The microscopic transport model UrQMD is used for this analysis. 
The predicted $\Lambda/\pi^-$, $\Xi^-/\pi^-$ and 
$\Omega^-/\pi^-$ ratios are enhanced due to rescattering in central Pb-Pb collisions 
as compared to peripheral Pb-Pb or p-Pb collisions. 
However, the enhancement is substantially smaller than observed experimentally.
The enhancement depends strongly on the kinematical cuts. The maximum enhancement is 
predicted around midrapidity. For $\Lambda$'s, strangeness {\it suppression}  
is predicted at projectile/target rapidity. 
For $\Omega$'s, the predicted enhancement can be as large as one order 
of magnitude. 
Comparisons of Pb-Pb data to proton induced asymmetric (p-A) collisions are  
hampered due to the predicted strong asymmetry in the various 
rapidity distributions of the different (strange) particle species. 
In p-Pb collisions, strangeness is locally (in rapidity) {\it not} conserved.  

The present comparison to the data of the WA97 and NA49 collaborations clearly 
supports the suggestion  
that conventional (free) hadronic scenarios are unable  
to describe the observed high (anti-)hyperon yields in central collisions. 
A reduction of the constituent
quark masses to the current quark masses $m_s \sim 230\,$MeV, $m_q \sim 10\,$MeV,
as motivated by chiral symmetry restoration,
yields hyperon production close to
the experimentally observed high values.
An ad hoc overall increase of the
color electric field strength (effective string tension of
$\kappa=3\,{\rm GeV/fm}$) yields similar results.
It has been suggested that these findings 
might be interpreted as a signal of a phase of nearly massless particles.
\end{abstract}


\newpage

\section{Introduction}

The investigation of strangeness production in relativistic heavy ion collisions 
is viewed as a powerful tool to study excited nuclear matter 
\cite{raf8286,koch86,koch88,bass98qgp,senger99,JPG,stock99,rafelski96,geiss98} and to search for 
the transition of (confined)  
hadronic matter to quark-gluon-matter (QGP). 
Several possible signals have been proposed for the observation of such a novel state 
of matter, but an unambiguous proof of its existence has not yet been achieved. 
For a recent review on the current status of possible QGP-signals see, e.g., 
\cite{bass98qgp}. 

Strange and especially multistrange particles are of interest:  
Their relative enhancement and the hadron ratios in central 
heavy ion collisions, with respect to peripheral (or proton induced) 
interactions, have been suggested as a signature for the transient existence of a 
QGP-phase \cite{raf8286,koch86,koch88}:  
The main argument is that the (chemical or flavour) equilibration times should be much 
shorter in the plasma phase than in 
a thermally equilibrated hadronic fireball of $T\sim 160\,$MeV.

The dominant production mechanism in an equilibrated (gluon rich) plasma phase,  
namely the production of $s\overline{s}$ pairs via gluon fusion 
($gg \rightarrow s \overline{s}$) \cite{raf8286},  
should allow for equilibration times similar to the 
interaction time of the colliding nuclei, and to the expected plasma lifetime 
(a few fm/c).

It has been shown recently also in microcanonical calculations 
\cite{bel98} that the equilibration times for kaons 
are similar to those of e.g. the pions, only in hadronic matter of very large 
energy density $e \gg e_0 = 0.15 \,{\rm fm}^{-3}$, where $e_0$ is  
the energy density of groundstate nuclear matter.

Recently, measurements by the WA97 and the NA49 collaborations 
clearly demonstrated the relative enhancement 
of the (anti-)hyperon yields ($\Lambda$, $\Xi$, $\Omega$) 
in Pb-Pb collisions as 
compared to p-Pb collisions \cite{and98a,and98,and99b,mar99,gab99,evans99}. 
The observed enhancement increases with the strangeness content ($|S|=1,2,3$) of 
the probe under investigation \cite{and98a,and98,and99b,mar99,gab99,evans99}. 
For the ($\Omega^- + \overline{\Omega^-}$)-yield the enhancement factor 
is as large as 15 !
The data cover the range in transverse energy $E_T$ where the 
''anomalous'' $J/\Psi$ suppression is observed, a different much sought after signal 
for deconfined matter.

Earlier experiments had also reported the enhanced production of 
strangeness (mostly kaons, see e.g. \cite{NA35,bari95,kinson95}).

Here, we investigate the 
strangeness enhancement within a microscopic transport model:  
hadronic and string degrees of freedom are employed in  
the Ultrarelativistic Quantum Molecular Dynamcis model (UrQMD) \cite{bas98,bleicher99}. 
The strange baryon yields for Pb$(158\,A\,{\rm GeV})$Pb collisions are computed  
vs. centrality and for p$(158\,{\rm GeV})$Pb collisions. 
The observed total yields of $\Lambda$'s, $\Xi$'s and $\Omega$'s are 
well described in the p-Pb case by the present model.
Strangeness enhancement is predicted in the present 
calculation for Pb-Pb due to rescattering. 
However, for central Pb-Pb collisions the 
experimentally observed hyperon yields are underestimated by 
the present calculations.
The discrepancy to the data increases strongly with the strangeness content of the hadron. 

An ad hoc overall increase of the color electric field strength (effective string tension of   
$\kappa=3\,{\rm GeV/fm}$), or, equivalently, a reduction of the constituent 
quark masses to the current quark masses enhances the hyperon yields to  
the experimentally observed high values.

Enhancement factors of $\approx 1.5 (2)$ for $\Lambda$'s, $\approx 2 (6)$ for $\Xi^-$'s, 
and $\approx 5 (13)$  for $\Omega^-$'s 
are obtained at midrapidity. The values in brackets are the results of the reduced masses/enhanced 
string tension calculations. 
The enhancement depends strongly on rapidity:  
for $\Lambda$'s, strangeness {\it suppression} is predicted at projectile/target rapidity. 
The hyperon yields depend strongly on rapidity and  are
asymmetric with respect to midrapidity for p-Pb reactions. 
Moreover, strangeness is not conserved locally in rapidity. 
Consequently, enhancement factors defined at midrapidity are not adequate for 
that comparison (p-Pb vs. Pb-Pb).

\section{The model}

UrQMD is a N-body transport model deviced to simulate heavy ion 
collisions in the laboratory energy range from several tens of MeV 
to several TeV per nucleon. 
A detailed documentation of the underlying concepts of the model and 
comparisons to experimental data are available in \cite{bas98,bleicher99}.
Binary elastic and inelastic collisions, many body resonance decays, as well 
as string decays are treated in the model. 
Inelastic collisions and decays are the only source for a change of  
the chemical composition. Elastic collisions change the momentum distributions 
of the hadrons only. String and resonance excitations are present both  
in primary nucleon nucleon as well as in secondary collisions. 
There are 55 baryon and 32 meson states (as well as their anti-particles) as 
discrete degrees of freedom in the model. 
Explicit isospin-projected states with masses up to 2.5 GeV are included. 
Strings can be populated as continuous degrees of freedom for masses $\ge 1.5\,$GeV. 
Experimental hadron cross sections and resonance decay widths are taken when available, 
otherwise the additive quark model is used to estimate the cross section.   
Hadrons produced through string decays are assigned a non-zero formation time which depends 
on the four-momentum of the particle. 
Newly produced particles cannot interact during their 
formation time. Leading hadrons may interact within their formation time with 
a reduced cross section proportional to the number of original valence quarks. 
The string tension $\kappa$ is set to $\kappa=1\,$GeV/fm \cite{LUND}.
The production probability of a $s\bar{s}$ pair is reduced as compared to
$u\bar{u} / d\bar{d}$-pairs in the string models \cite{LUND} according to the
Schwinger formula \cite{schwing51}
\begin{equation}
\gamma_s=\frac{P(s\bar{s})}{P(q\bar{q})}=\exp 
\left(- \frac{\pi (m_s^2-m_q^2)}{2\kappa}\right)\,.
\label{eq1}
\end{equation}

In central high energy heavy ion collisions the 
string density can be so high
that the color flux tubes overlap \cite{biro84,sor92}.
Consequently, the superposition of the color electric fields
yields an enhanced particle production \cite{biro84,sor92}. 
In particular, the heavy flavors and diquarks are dramatically
enhanced due to a higher effective string tension \cite{sor92,gyulassy90,gerland95}.
As a consequence the string fragmentation probability into hyperons is enhanced as well. 
The increase of the string tension from $\kappa = 1\,$GeV/fm to $\kappa = 3\,$GeV/fm
enhances $\gamma_s$ from
$\gamma_s \approx 0.3$ to $\gamma_s \approx 0.65$
(for constituent quark masses of $m_q=0.3\,$GeV and $m_s=0.5\,$GeV).
However, as seen from eq.(\ref{eq1}), an increase of the string tension $\kappa$ 
is equivalent to a decrease of the difference of the squared constituent 
quark masses. Therefor, this enhancement could also be due to a drastic drop 
in the masses due to chiral symmetry restoration \cite{brownrho,papazoglou}. 
If we assume a reduction to the current quark masses, $m_s \sim 230\,$MeV, $m_q \sim 10\,$MeV, 
the strangeness to nonstrange suppression factor is also increased 
to $\gamma_s \approx 0.65$. 
Hence, such doubling of $\gamma_s$ might be interpreted as a signal of a phase 
of nearly massless particles.  
Here we increase the ratio overall, equivalent to an increase of $\kappa$ from 
$1\,{\rm GeV/fm}$ to $3 \,{\rm GeV/fm}$.
Total energy is conserved in this process. Hence, the production    
of non-strange hadrons is only moderately modified -    
the pion number changes by less than $\approx  4\%$. 
These $\gamma_s$-values are not to be mixed up with (although striking similar to) 
the Becattini-values for $e^+e^-$ ''thermal'' string analysis and for the  
heavy ion analysis \cite{becattini98,yen99,rafelski91}

\section{Results and Discussion}

The yields of strange baryons per event are shown in Fig.1 as a function of the number 
of participants $N_{\rm part}$ for the collisions Pb$(158\,A\,{\rm GeV})$Pb and 
p$(158\,{\rm GeV})$Pb. 
The $\Lambda+\overline{\Lambda}$- (circles), 
$\Xi^- + \overline{\Xi^-}$- (squares), and $\Omega^- + \overline{\Omega^-}$- (triangles) values 
are depicted, their strangeness content is $|S|=1,2\,$and$\,3$, respectively. 
A midrapidity cut $|y-y_{\rm cm}|<0.5$ has been applied in accord with 
\cite{and98a}.
The stars correspond to experimental data of the 
WA97 collaboration \cite{and98a}. Open symbols represent the results 
of the standard UrQMD calculations. 
Full symbols exhibit the UrQMD calculations 
with the decreased mass square difference, or, equivalently, the 
enhanced string tension $\kappa=3\,$GeV/fm, 
for the most central collisions ($N_{\rm part}\ge 300$).

The hyperon yields increase nearly like a power-law
$\rm{ln}(Y/{\rm event})={\rm ln}(N_{\rm part}^{\alpha})$ with the exponent
$\alpha \approx 1.1$ for $\Lambda$'s and $\Xi$'s and $\alpha \approx 2$ for $\Omega$'s.  

All baryon yields increase continuously with centrality up to 
very central events. 

The l.h.s. points ($\langle N_{\rm part} \rangle \approx 4$) are
the inclusive p($158\,{\rm GeV}$)Pb yields. 
The data agree with the calculations for 
$\Lambda$'s, $\Xi$'s, and $\Omega$'s in p-Pb collisions.
The UrQMD results for Pb-Pb, however 
are systematically below the experimental data (stars) for $N_{\rm part} >  100$. 
Small deviations for the $\Lambda$'s (open circles) give way to a larger 
discrepancy ($\approx$ factor 2)
for the $\Xi$'s (open squares).
The $\Omega$'s (open triangles) are underestimated by more than an order 
of magnitude in the UrQMD calculations. 

The yields as calculated with decreased mass square difference, or, equivalently, with 
the higher string tension ($\kappa=3\,{\rm GeV/fm}$) (full symbols)
are systematically above the UrQMD calculations ($\kappa=1\,{\rm GeV/fm}$). 
The apparent improvement of the comparison between data and model calculations 
should not be viewed as a solution to the puzzling enhancement factors: 
In particular, the nearly constant displacement factor of the data to the 
model indicates a different mechanism (the mass reduction or $\kappa$ increase 
must physically depend on the centrality). 
The surprising complete agreement for $\Lambda$'s and $\Xi$'s is not understood. 
The $(\Xi^- + \bar{\Xi^-})/(\Lambda + \bar{\Lambda})$  ratio is calculated as
$0.07\pm 0.01$ and $0.12 \pm 0.01$ in the UrQMD and the decreased masses/enhanced string
tension calculations, respectively, with the latter being in accord with  the
experimental values of the WA97 ($0.11\pm 0.003$) \cite{and99b} and NA49 ($0.13\pm0.04$) 
\cite{mar99} collaborations.
The small deviation of the $\Omega$ yield can be due to 
the slightly underestimated p-Pb data point. 
The mean values for the $\Omega$ yield differ by a factor $\approx 1.6$. 
As a consequence the $\Omega/\Xi$ ratio is predicted to be $\approx 0.12\pm0.02$ 
in contrast to the WA97 value of about 0.2 \cite{and99b}.
This discrepancy was also found in \cite{rafelski99}, where the $\Omega$ yield 
could also not be consistently described with other particle yields in the 
framework of a Fermi statistical model analysis, indicating the need for 
an additional production mechanism. 
Note also our discussion of the 
uncertainties in the centrality variable $N_{\rm part}$.

In the calculations $N_{\rm part}$ is determined via the 
scaled number of $\pi^-$'s (in $4 \pi$ geometry), $N_{\rm part} \approx 0.5 \langle \pi^- \rangle$. 
This number was found to be a robust observable, since it is directly proportional to 
the overlap volume of the colliding nuclei and thus to the 
desired quantity $N_{\rm part}$. 
One may also determine $N_{\rm part}$ by counting all collided baryons.
This quantity is not strictly proportional to the geometrical   
overlap volume and therefore yields a different centrality criterion.
Various experimental $N_{\rm part}$-determinations have been using 
Glauber model estimates, the wounded nucleon model, or the Venus model as well 
as Fireball-extrapolations from limited acceptance data. 

Strange meson yields are enhanced in the decreased mass square difference/higher
string tension calculations if compared to the UrQMD calculations.   
The factors are $\approx 1.5$ for kaons and $\approx 2.7$ for $\phi$'s, similar
to the $\Lambda$ and $\Xi$ with the same number of strange quarks.
The comparison of the ratios K$^+/\pi$, K$^-/\pi$ to
preliminary NA49 data \cite{sikler99} shows that the agreement improves for  
the enhanced $\gamma_s$ calculations, whereas the $\phi/\pi$ ratio
seems to be overestimated by a factor $\approx 1.6$ (see Table 1).
Note, that the $\phi$ measurements by the NA50 collaboration \cite{willis99} indicate
significantly higher yields than those obtained by NA49. Additionally, the
inverse slope parameters are different ($220\,$MeV and $290\,$MeV).

Can one observe an anomalous, sudden threshold enhancement of the (anti-)hyperon 
yields with increasing 
centrality (i.e. number of participants) ? 
The hyperon-to-pion ratios are shown in fig.~2 as a function of centrality. 
The centrality variable chosen here is the number $\langle \pi^- \rangle$ in $4\pi$ geometry.  
The pion number $\langle \pi^- \rangle$ scales with the geometrical overlap of the 
two nuclei. 
 
A linear increase of the hyperon yields with $\langle \pi^- \rangle$ 
will yield constant particle ratios $Y/\pi$. 
However, these ratios $Y/\pi^-$ increase with $\langle \pi^- \rangle$ as can be 
seen in Fig. 2. These ratios are for the full yields in 4$\pi$ geometry. 
The increase in the ratios is more pronounced if rapidity/$p_t$ cuts are applied \cite{soff99}. 

The full symbols in Fig.2 correspond to the reduced mass/increased $\kappa=3\,{\rm GeV/fm}$ 
calculations. 
Open symbols are the UrQMD calculations ($\kappa=1\,{\rm GeV/fm}$). 
There is a clear enhancement of the hyperon-to-pion ratio in central Pb-Pb 
collisions as compared 
to the inclusive p-Pb and peripheral Pb-Pb collisions. 
The predicted threshold enhancement of both $\Xi$'s and $\Omega$'s 
is between $N_{\rm part} \approx 10$ and 25. 
This indicates that multiple collisions and rescattering effects become important 
and thus allow for multiple production of heavy strange quarks \cite{mat89}.
For larger $\langle \pi^- \rangle$-values the ratios increase only moderately. 
The WA97 collaboration has reported a nearly linear increase in the yields for 
$N_{\rm part} > 100$.

If constituent quark mass reduction or collective string effects ($\kappa=3\,$GeV/fm) 
are taken into account for the 
more central collisions (full symbols) then the overall enhancement of strange particle 
production becomes also stronger. 
The change of the ratio $Y/\pi^-$  due to
the reduced mass/increased string tension grows with the strangeness content of the probe.
The star in Fig.2 represents an estimate for the $\Xi^-/\pi^-$ ratio in 4$\pi$ geometry 
of the NA49 collaboration \cite{sikler99}.
Again, it coincides with the reduced mass/enhanced string tension 
calculations. 

The strangeness enhancement factor 
\begin{equation}
E_Y = (Y/\pi¯)_{\rm Pb-Pb, central} / (Y/\pi¯)_{\rm Pb-Pb, peripheral} \nonumber 
\end{equation}
is predicted in Fig.3 as a function of rapidity for $\Lambda$, $\Xi^-$ and $\Omega^-$ 
hyperons, respectively. 
The full range of transverse momenta is taken $p_t \ge 0\,$GeV/c.
The open symbols correspond to the 
UrQMD calculations ($\kappa=1\,$GeV/fm).  
Full symbols are the enhancement factors if the reduced mass/enhanced string tension 
calculations are taken into account for the central collisions. 
The enhancement factors reach a maximum around midrapidity
and decrease continuously towards target/projectile rapidity, thus 
demonstrating the importance of secondary collisions. 
Values of $E_{\Lambda} \approx 1.5$, $E^*_{\Lambda} \approx 2.0$, $E_{\Xi^-} \approx 2.2  $, 
$E^*_{\Xi^-} \approx 6.0 $, 
$E_{\Omega^-} \approx 4.8$, and $E^*_{\Omega^-} \approx 13 $ are obtained at midrapidity.  
The stars indicate the reduced mass/enhanced string tension calculations  
for the central events. 
The enhancement factors grow with the
strangeness content of the particle.   
This is in line with the experimental finding \cite{and98,and98a}.  
Strangeness {\it suppression} is predicted for 
the $\Lambda$'s at target/projectile rapidity. 
This is due to the associated production of $\Lambda$'s via e.g. $pp\rightarrow p\Lambda K^+$, 
where the produced $\Lambda$ carries the full momentum of the incident proton. This occurs more 
frequently 
in peripheral than in central Pb-Pb collisions.

The enhancement factors $E_Y$ here are not determined from the ratios of 
central Pb-Pb to p-Pb collisions:  
the rapidity distributions in p-Pb are strongly asymmetric (see fig.4). 
The anti-baryons ($\overline{p}, \overline{\Lambda}, \overline{\Xi^-},$ etc.) are 
predominantly produced at midrapidity, while the distributions of 
$K$'s, $\Lambda$'s and $\Xi^-$'s (and $\pi$'s)  
are strongly shifted to target rapidity and are additionally asymmetric. Thus, a comparison 
of midrapidity yields in central Pb-Pb and p-Pb is not adequate.  

The asymmetry of the collision system p-Pb is also demonstrated by the 
\mbox{(net-)strangeness} rapidity distribution $dS/dy$ in Fig.5. 
$s$-quarks (S=-1) (squares) and $\bar{s}$-quarks (S=1) (circles) and their sum are 
depicted. The $S=1$ values represent the sum of 
$(K^+ + K^0 + \bar{\Lambda}+ \bar{\Sigma}) + 2\,(\bar{\Xi}) +3\,(\bar{\Omega})$, 
while the $S=-1$ values are determined by 
$(K^- + \bar{K^0} + \Lambda+ \Sigma) + 2\,(\Xi) +3\,(\Omega)$.    
The target rapidity region is dominated by  strangeness production, while the anti-strangeness 
dominates around midrapidity. 
If confirmed experimentally this result is important
for the production of (multi-)hypernuclei or strangelets: The finite 
initial net strangeness will support the 
strangeness distillery mechanism \cite{cgreiner}. 

\section{Conclusions}

(Anti-)Hyperon yields in central Pb($158\,A\,$GeV)Pb collisions 
\cite{and98a,and98,and99b,mar99,gab99,evans99} are 
strongly enhanced in comparison  
to p($158\,$GeV)Pb and peripheral Pb-Pb collisions. 
This enhancement grows quadratically with the strangeness content of the hyperon. 
The present model predicts that strangeness enhancement occurs as a threshold effect 
already at rather small number 
of participants $(< 25)$ due to rescattering. 
For larger participant numbers the yields grow almost linearly with $N_{\rm part}$. 
Reducing the effective masses of the constituent quarks to the current quark mass values 
(or, equivalently, increasing the  string tension) for central collisions 
yields large additional enhancement, which grows 
with the strangeness content in a similar manner as the data. 
The experimentally observed high hyperon yields in 
central Pb-Pb collisions cannot be predicted in our conventional  hadron-string dynamics.  
The rapidity distributions in p-Pb collisions are asymmetric with respect to midrapidity.  
Thus, comparing the midrapidity p-A yields to peripheral or central Pb-Pb-data 
is misleading. Strangeness is locally (in rapidity) not conserved in p-Pb collisions.

The comparison to data of the WA97 and NA49 collaborations shows clearly 
that there seems to be no conventional hadronic scenario which is able to 
describe the high hyperon yields. 
This is so far the only clear signal which contradicts transport calculations 
based on hadronic and string degrees of freedom and therefore indicates an  
extraordinary behaviour of hot and dense matter in the early phase, possibly 
a transition to a chirally restored state.

\pagebreak
\begin{table}
\caption{K$^+/(0.5(\pi^++\pi^-))$,K$^-/(0.5(\pi^++\pi^-))$, and $\phi/(0.5(\pi^++\pi^-))$ 
ratios in central Pb$(158\,A\,{\rm GeV})$Pb collisions as calculated with 
UrQMD (left column), with decreased masses/enhanced string tension (see text, middle column) and 
compared to preliminary NA49 data (right column).}
\begin{center}   
\begin{tabular}{|r||l|l|r|}
\hline
           & UrQMD & mod. UrQMD, $\gamma_s$ & prel. NA49 \\
\hline
K$^+/\pi$   & $ 0.12 \pm 0.01 $  &  $0.19    \pm 0.01$    & $0.16\pm 0.013$ \\
K$^-/\pi$   & $0.065  \pm 0.01$ &  $0.1 \pm 0.01    $    & $0.09 \pm 0.008$ \\
 $\phi/\pi$ & $0.007 \pm 0.001$  & $ 0.02 \pm 0.001  $    & $0.0127 \pm 0.0014 $\\
\hline
\end{tabular}
\end{center}
\end{table}

\newpage


\begin{figure}
\caption{(Anti-)Hyperons per event at midrapidity $|y-y_{\rm cm}|< 0.5$ as a function 
of the number of participants $N_{\rm part}$ for Pb-Pb and p-Pb collisions at 
$158\,A\,$GeV. 
The yields of $\Lambda+\overline{\Lambda}$ (circles), 
$\Xi^- + \overline{\Xi^-}$ (squares) 
and $\Omega^- + \overline{\Omega^-}$ (triangles) 
increase continuously with $N_{\rm part}$. Stars are 
experimental data by the WA97 collaboration 
(for $\Lambda$'s, $\Xi$'s and $\Omega$'s). 
Open symbols represent the UrQMD calculations. Full symbols are the results 
of the reduced masses/enhanced string tension calculations.} 
\end{figure} 

\begin{figure}
\caption{Hyperon to $\pi^-$ ratios in p-Pb and Pb-Pb at $158\,A\,$GeV 
as a function of the number 
of negativly charged pions (in 4$\pi$ geometry). 
Open symbols are UrQMD calculations, 
while full symbols at the more central collisions 
($N_{\pi^-} \ge 300$) are obtained from the reduced masses/$\kappa=3\,{\rm GeV/fm}$ calculations. 
All ratios increase towards more central collisions indicating 
the strangeness enhancement which grows with the strangeness content of the 
particle. The star is an estimate of the $\Xi^-/\pi^-$ ratio in central collisions by the 
NA49 collaboration.}  
\end{figure}

\begin{figure}
\caption{Rapidity dependence of the strangeness enhancement factors 
$E_{\Lambda}$, $E_{\Xi}$ and $E_{\Omega}$, defined 
as the relative enhancement of the hyperon yields in the most central compared 
to peripheral Pb-Pb collisions in the respective rapidity bin. 
The reduced masses/($\kappa=3\,{\rm GeV/fm}$) calculations (full symbols) yield 
an additional enhancement as compared to the standard UrQMD calculations.
The enhancement is maximum at midrapidity. For the $\Lambda$'s, 
strangeness suppression is predicted at target/projectile rapidity.}
\end{figure}

\begin{figure}
\caption{Rapidity distributions of pions, kaons, anti-protons and (anti-)hyperons in 
p($158\,$GeV)Pb collisions. While anti-baryons are predominantly 
produced at midrapidity, the other distributions are strongly asymmetric with 
respect to midrapidity ($y_{\rm cm}=0$), shifted towards target rapidity.}
\end{figure}

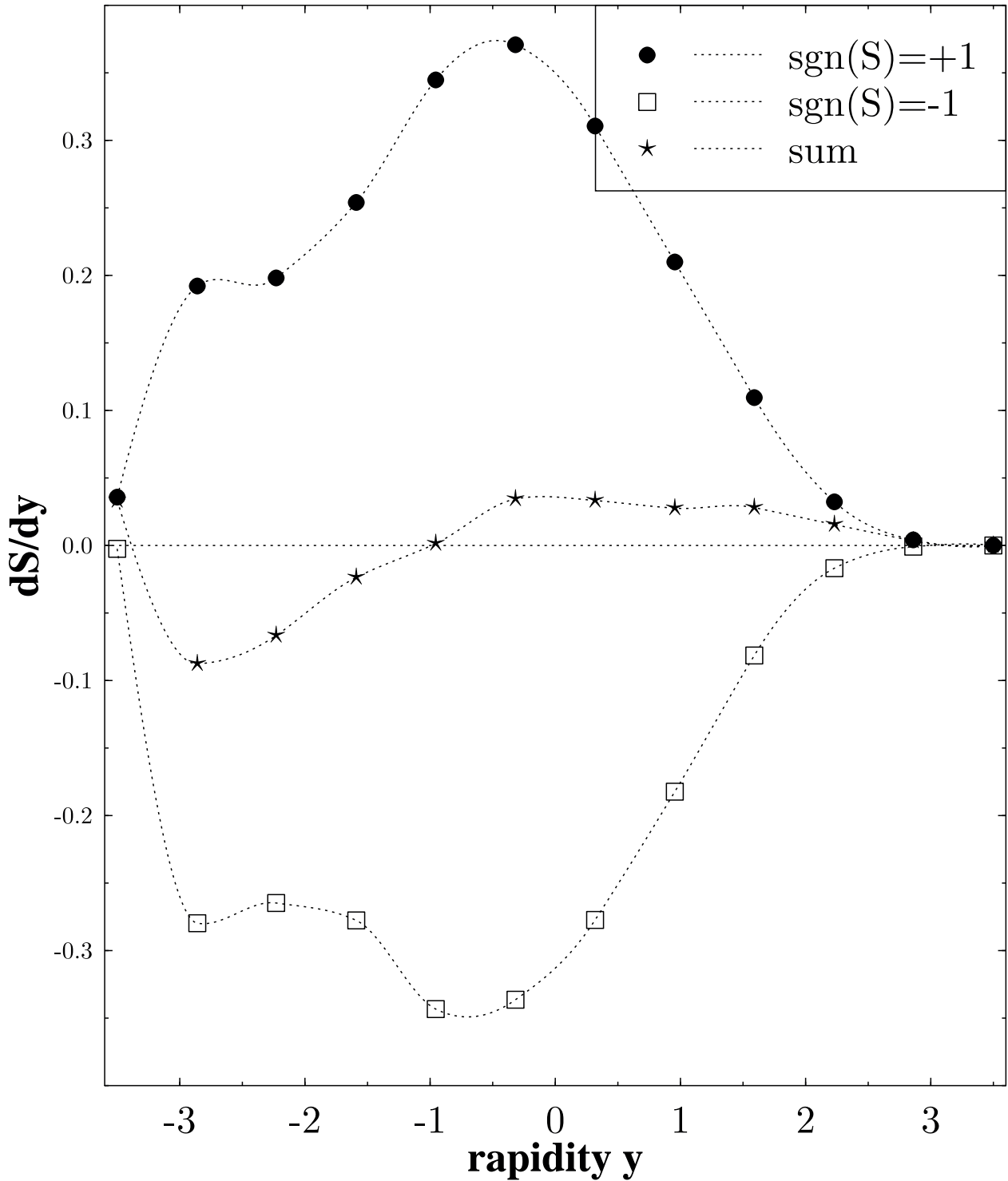
\begin{figure}
\caption{Strangeness rapidity distributions dS/dy in p$(158\,$GeV)Pb collisions 
of positive strangeness ($S=1$) (circles), 
negative strangeness ($S=-1$) (squares), and the sum of both (stars), respectively. 
Strangeness is locally not conserved.}
\end{figure}

\newpage
\pagestyle{empty}
\begin{figure}[t]
\centerline{\psfig{figure=letterfig1_newest.eps}}
\end{figure}

\newpage
\samepage{
\begin{figure}[h]
\centerline{\psfig{figure=letterfig2_newest.eps}}
\end{figure}
}
\newpage

\begin{figure}[h]
\centerline{\psfig{figure=letterfig3_newest.eps}}
\end{figure}

\newpage

\begin{figure}[h]
\centerline{\psfig{figure=letterfig4_newest.eps}}
\end{figure}

\newpage

\begin{figure}[h]
\centerline{\psfig{figure=letterfig5_newest.eps}}
\end{figure}

\enddocument